\documentclass[aps,prl,twocolumn,superscriptaddress,showpacs]{revtex4}
\usepackage{graphicx}

\begin{document}

\title{A cascade electronic refrigerator using superconducting tunnel junctions}

\author{H. Q. Nguyen}
\affiliation{Low Temperature Laboratory, Department of Applied Physics, Aalto University School of Science, P.O. Box 13500, 00076 Aalto, Finland}
\affiliation{Nano and Energy Center, Hanoi University of Science, VNU, Hanoi, Vietnam}
\author{J. T. Peltonen}
\affiliation{Low Temperature Laboratory, Department of Applied Physics, Aalto University School of Science, P.O. Box 13500, 00076 Aalto, Finland}
\author{M. Meschke}
\affiliation{Low Temperature Laboratory, Department of Applied Physics, Aalto University School of Science, P.O. Box 13500, 00076 Aalto, Finland}
\author{J. P. Pekola}
\affiliation{Low Temperature Laboratory, Department of Applied Physics, Aalto University School of Science, P.O. Box 13500, 00076 Aalto, Finland}

\begin{abstract}
	Micro-refrigerators that operate in the sub-kelvin regime are a key device in quantum technology. A well-studied candidate, an electronic cooler using Normal metal - Insulator - Superconductor (NIS) tunnel junctions offers substantial performance and power. However, its superconducting electrodes are severely overheated due to exponential suppression of their thermal conductance towards low temperatures, and the cooler performs unsatisfactorily - especially in powerful devices needed for practical applications. We employ a second NIS cooling stage to thermalize the hot superconductor at the backside of the main NIS cooler. Not only providing a lower bath temperature, the second stage cooler actively evacuates quasiparticles out of the hot superconductor, especially in the low temperature limit. The NIS cooler approaches its ideal theoretical expectations without compromising cooling power. This cascade design can also be employed to manage excess heat in other cryo-electronic devices.
\end{abstract}
\maketitle

Growing out of nanotechnology and low temperature physics, ultra-sensitive microdevices employing quantum physics initiate technologies that were inaccessible previously. For example, bolometers \cite{IrwinAPL, Enss} in spaceborne telescopes \cite{Tauber,Herschel,Hitomi} require a dilution cryostat reaching sub 100 mK temperatures to extend our observation into the deep space, or quantum bits \cite{Schumacher,Martinis,Johnson} enable in principle an exponential enhancement of computing power and probably could simulate nature. The main unit of these devices, a superconducting element that takes on quantum physics, requires cryogenic temperatures that are typically provided by a dilution cryostat. This bulky, complicated, and expensive machine inspires a search for alternative approaches. 

In the sub-kelvin regime, there are multiple choices of thermal machines that explore new physics at the microscale. They include quantum dots \cite{Prance}, single ions \cite{Singer}, microelectromechanical systems \cite{Richard}, piezoresistive elements \cite{vanBeek}, or cold atoms \cite{Esslinger}. Although promising, most of these machines are premature or only at the stage of conceptual designs. The most practical one, a cooler using a tunnel junction between a normal metallic and a superconducting electrode, \cite{NahumAPL, LeivoAPL96, MuhonenRPP,GiazottoRMP06} offers substantial electronic cooling power. Under a bias near the superconducting energy gap, hot electrons tunnel out of the normal metal into the superconductor. This cools down electrons in the normal metal, but at the same time deposits a considerable amount of heat into the superconducting lead. Suppressing hot quasiparticles is hard, as the superconducting gap prevents them from thermalizing with the phonon bath. This is a serious problem in powerful devices, where a large current constantly generates quasiparticles that overheat the superconductor severely. This heat also flows back to the normal metal as $P = \beta IV$, $IV$ is the total input power, and $\beta$ is a phenomenological coefficient in the range of 1-5\% \cite{ONeilPRB12}. To date, the most efficient method is to employ a normal metal as a quasiparticle trap \cite{CourtPRB08, VasenkoPRB10, SukumarPRB12, Luukanen, ONeilPRB12, Agulo04, VoutilainenPRB,PekolaAPL00}, and place it as close as possible to the overheated region of the superconductor.

\begin{figure*}[t]
\begin{center}
\includegraphics[width=6.5in,keepaspectratio]{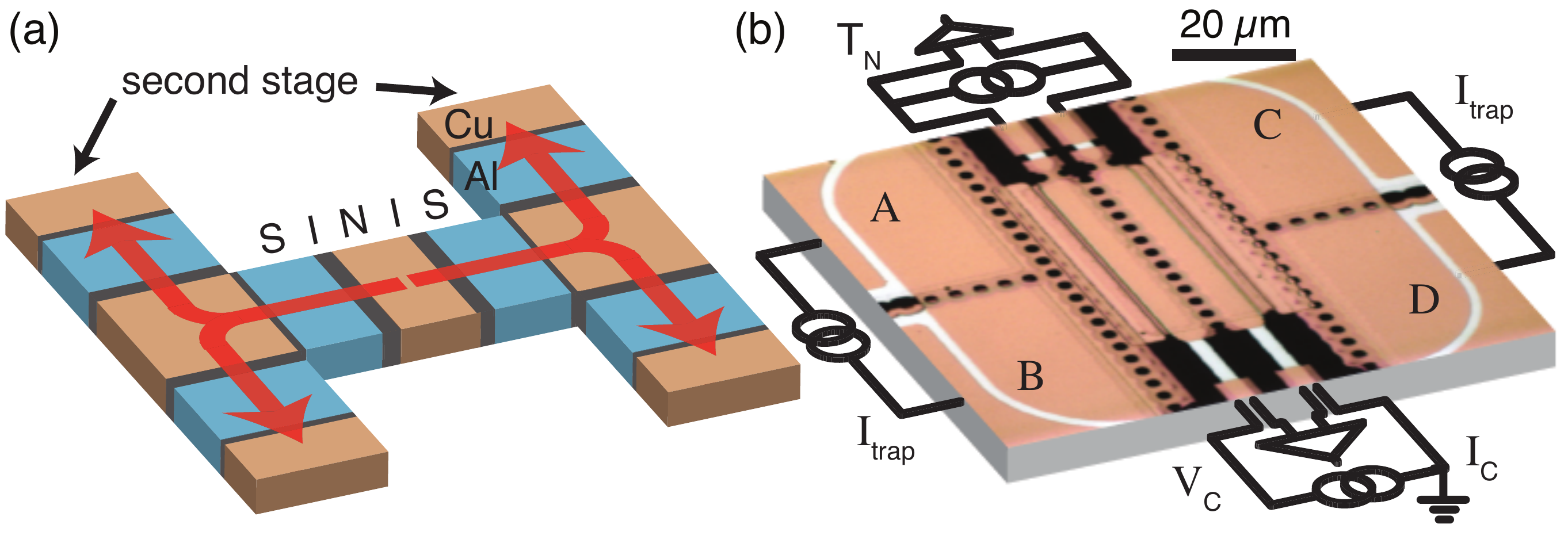}
\caption{\textbf{A two-stage cooler:} (a) Working principle: back sides of the main cooler is connected to two other SINIS coolers. These second stage coolers act as "active quasiparticle traps" and help to thermalize the hot superconductor of the main device. Red arrows represent heat flows out of the central island. (b) Optical image of the actual device and its measurement setup: each of the holes which suspend the Cu island on top of Al electrodes, has a diameter of 2 $\mu$m. Each of the two main cooler junctions are sized 50 $\times$ 4 $\mu$m$^2$ and are biased with $I_C$. The two secondary coolers (A-B and C-D) are located 3 $\mu$m away from the main junction. The second-stage coolers are biased with floating currents $I_{\text{trap}}$ from battery-powered sources. The pair of small NIS junctions is an electron thermometer that probes temperature $T_N$ of the normal island.}
\label{diagram5}
\end{center}
\end{figure*}

A bi-polar device, here a thermal machine with a hot and a cold end, can be operated in a multi-stage setup. The cold part provides cooling for the hot region from a previous stage in a cascade architecture. This strategy is best seen in the design of common cryostats \cite{Lounasmaa}, where, the ultimate temperature below 10 mK is only reached by a series combination of several cooling stages, beginning from those at 4 K having cooling powers on the order of watts, at least one intermediate stage cooled to about 500 mK by evaporating $^3$He with about 100 mW cooling power till finally the phase separation between liquid $^3$He and liquid $^4$He decreases the entropy of the system and cools it to temperatures below 10 mK with few $\mu$W of cooling power. Such a cascade architecture is widely employed also in various practical systems, such as thermoelectric coolers employing Peltier effect \cite{HwangPeltier}, and outside refrigeration technology in vacuum pumps and air handling units. However, it has not been experimentally realized with solid-state NIS coolers \cite{CamarasaG} where the overheated superconductors are the main issue. 

In this work, we realize such a concept, as seen in Fig. 1, and prove that the double stage NIS refrigerator performance is significantly improved. Based on an established technique \cite{NguyenAPL}, a powerful large-junction SINIS cooler (two NIS junctions back-to-back) is fabricated with a well thermalized superconductor of aluminum, boosted by an advanced quasiparticle trapping strategy. A normal metal layer is engineered directly underneath the hot superconductor to improve thermalization. Separated by a thin oxide layer, the superconductor is not affected by inverse proximity effect, while quasiparticles dissipate into the normal metal to lose energy with phonons \cite{NguyenNJP}. To date, this design yields the best performance of the SINIS cooler \cite{NguyenPRN}. Here, we extend this quasiparticle drain and directly connect it to the second stage cooler. During operation, the center SINIS cooler generates quasiparticles in the two superconducting electrodes. This heat dissipates in the normal metal drain and is actively evacuated by the "active quasiparticle trap".

An actual image of the device together with a sketch of the measurement setup is shown in Fig. 1 (b). The fabrication starts with sputtering 100 nm of Cu on a Si wafer and patterned it into quasiparticle drains using photolithography and chemical wet-etch. The wafer is loaded into the sputter again, and deposit 10 nm Al, followed by a slight oxidation in a mixture of Ar:O$_2$ at ratio 1:10, at the pressure of 2 mbar for 2 minutes. This thin AlOx layer prevents the inverse-proximity effect in the superconductor. Then, 200 nm of Al is sputtered and oxidized at 7 mbar in pure O$_2$ for 5 minutes, followed by a deposition of 60 nm of Cu. The device is patterned by photolithography, which aligns the junction to the drain. This step is crucial as a small misalignment can short the drain and the top Cu layer along the hole array. The metals are wet-etched as in Ref. \cite{NguyenAPL}. A third photolithography and Cu wet-etch define the NIS junctions with the size of the main cooler of 50 $\times$ 4 $\mu$m$^2$. The active trap with junction area 25$\times$20 $\mu$m$^2$ is 10 $\mu$m away from the hole array of the main cooler. In the last photolithography step, a trench in the resist is opened, followed by a deposition of 100 nm Cu and a lift-off so that the new Cu connects the drain of the main cooler to the top normal metal of the active trap. 

The sample is measured in a dilution cryostat using standard current-voltage measurements in a four-probe configuration. To ensure no interferences with the operation of the main cooler, the traps are biased by floating current sources using a standard 1.5 V AA battery and a large series resistor. This way, a set of measurements with discrete trap currents is collected by changing the resistor in series. Temperature of the normal metal is monitored with a pair of smaller NIS junctions, which are also floating and biased at 7.5 nA. Voltage drop on these NIS junctions are calibrated against the cryostat temperature when the traps and the main cooler are off \cite{NguyenPRN}.

\begin{figure}[t]
\begin{center}
\includegraphics[width=3in,keepaspectratio]{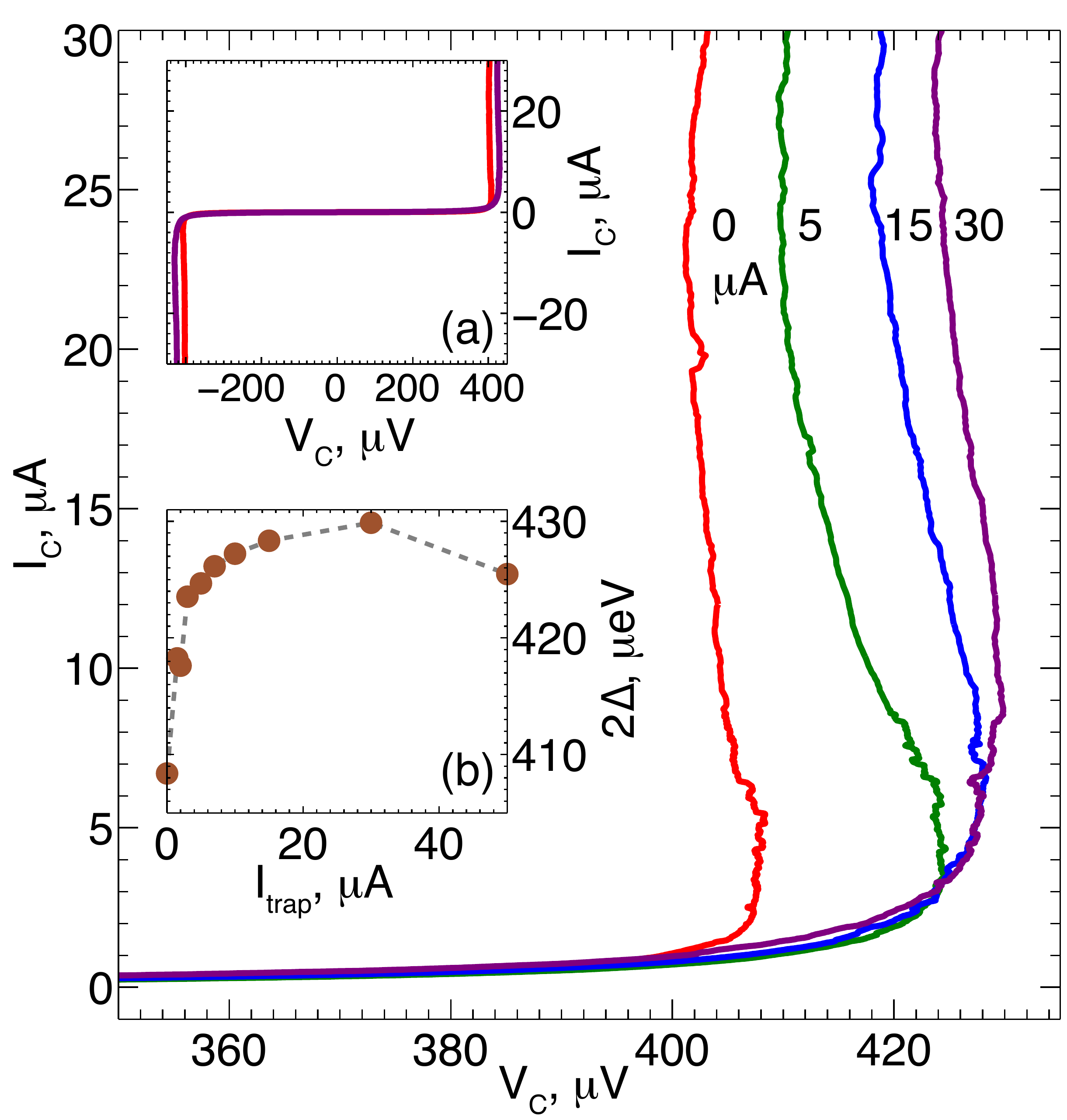}
\caption{\textbf{Transport measurement.} Main figure: close up view of the current voltage characteristics (IV curves) of the main cooler for different values of the trap current: $I_{\text{trap}}$ = 0, 5, 15, 30 $\mu$A, respectively, at the bath temperature of 70 mK. The numbers refer to the corresponding values of $I_{\text{trap}}$. Inset (a), zoomed-out IV curves at 70 mK for $I_{\text{trap}} = 0$ and 30 $\mu$A. Inset (b), value of the energy gap of the central cooler defined as $\Delta =$ max$(eV_C, I_C < 10$ $ \mu$A$)$ as a function of $I_{\text{trap}}$. The dashed line connects the dots as a guide for eyes.}
\label{IVs}
\end{center}
\end{figure}

The thermalization effect from the active trap is immediately demonstrated by the reduced overheating and consequently increased superconducting gap value of the main cooler at finite values of $I_{\text{trap}}$, as shown by the current voltage characteristics in Fig. 2. We take the superconducting gap $\Delta$ to be presented by the maximal value of the voltage in the range $I_C < 10$ $\mu$A. With this practical definition we obtain $2\Delta = 408, 425, 428,$ and $430$ $\mu$eV when the current through the secondary coolers obtains values $I_{\text{trap}}$ = 0, 5, 15, and 30 $\mu$A, respectively. At low temperatures, the superconducting gap follows $\Delta(T) \approx \sqrt{2\pi k_B T_S/\Delta_0} e^{-\Delta_0/k_B T_S}\Delta_0$. For $2\Delta_0$ = 430 $\mu$eV, this yields a superconductor temperature $T_S$ = 700 mK when the trap is off. As the input current on the main cooler increases, electrons in the superconductor are overheated as the active trap cannot cope with the flux of extra quasiparticles, and temperature increases so much that the value of the gap decreases. This is seen as back bending of the IV curves, especially for small trap currents around $I_{\text{trap}}$ = 5 $\mu$A. The effect of the active trap is summarized in Fig. 2 (b): the superconducting gap is maximized at an optimal value of $I_{\text{trap}}\sim$ 30 $\mu$A. Above this value the gap shrinks again due to the overheating of the active trap. Our result demonstrates clearly that the superconductor temperature can be lowered by applying a finite trap current $I_{\text{trap}}$.

\begin{figure*}[t]
\begin{center}
\includegraphics[width=5in,keepaspectratio]{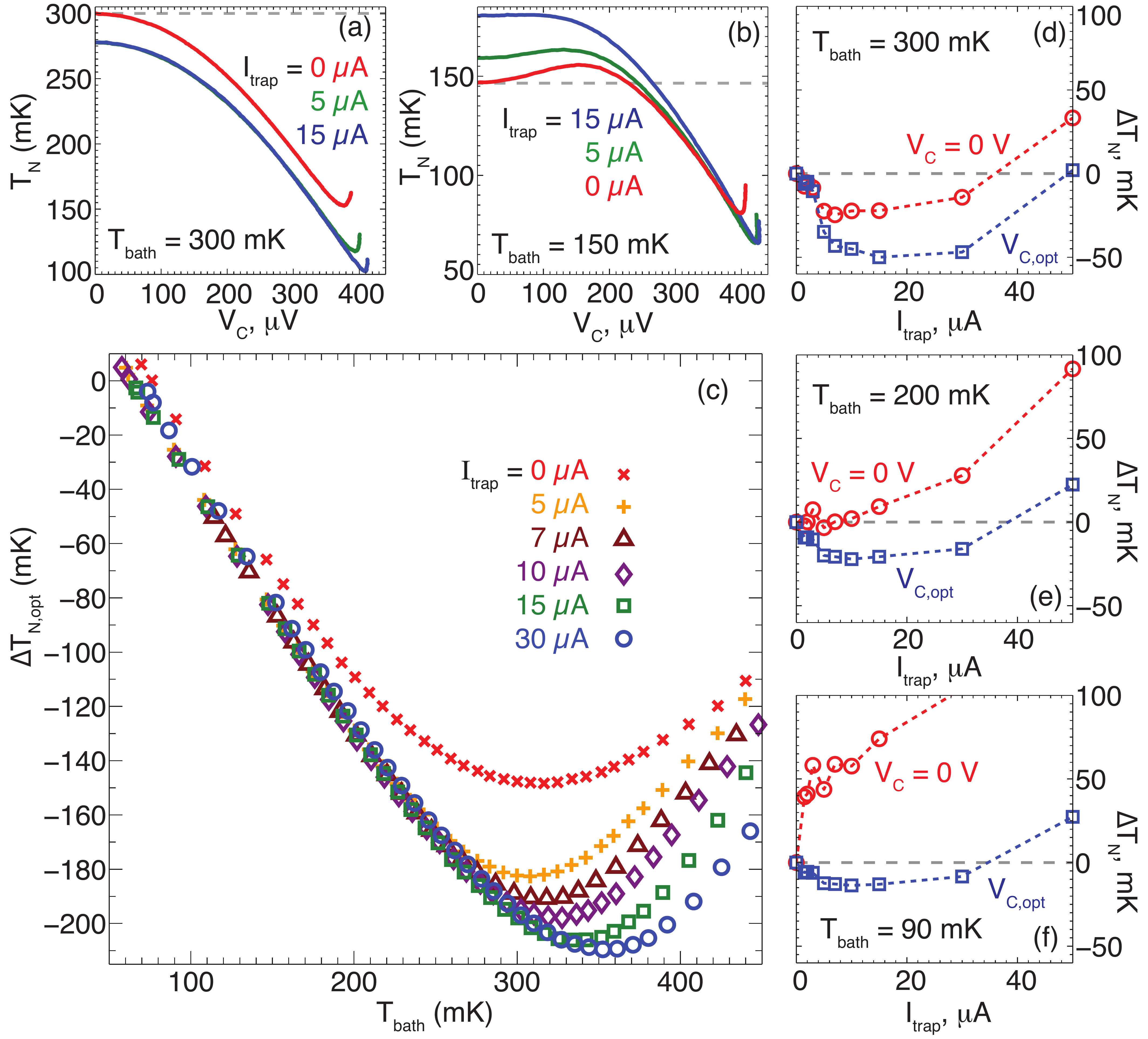}
\caption{\textbf{Performance of the double stage cooler}. (a) Electronic temperature of the central normal metal island $T_N$ as a function of voltage across the cooler $V_C$ at bath temperature 300 mK for $I_{\text{trap}}$ = 0, 5, and 15 $\mu$A in (a), and for the same trap currents at 150 mK in (b). Dashed lines represent bath temperature. (c) Optimum cooling $\Delta T_{\text{N,opt}} = T_{\text{bath}}-T_{\text{N,opt}}$ as a function of bath temperature at various values of $I_{\text{trap}}$. (d) Electronic temperature differences due to working trap cooler $\Delta T_N=T_N(I_{\text{trap}})-T_N(I_{\text{trap}}=0)$ measured on the normal metal central island at bath temperature 300 mK as a function of $I_{\text{trap}}$ when the cooler is off ($V_C = 0$ V, red circles) and at optimum bias point ($V_{\text{C,opt}}$, blue squares). (e) Similar data to (d) at the bath temperature of 200 mK, and in (f) at 90 mK.}
\label{cooling}
\end{center}
\end{figure*}

Figure 3(a) captures the basic behavior of the device at the bath temperature of 300 mK for three representative trap currents: $I_{\text{trap}}$ = 0, 5, and 15 $\mu$A. The temperature of the normal island is lowered from 300 mK to 275 mK when the central cooler is off, $V_C =0$,  by the cooling of the active traps. Due to the cooled superconducting electrodes, the central device now operates at a lower bath temperature. It extends the achievable temperature to 115 mK, whereas it is seen to be 150 mK at $I_{\text{trap}} = 0$. A higher trap current, $I_{\text{trap}}$ = 15 $\mu$A, improves the performance further, reaching a base temperature of 100 mK, which is a state-of-the-art figure for such a high-power cooler. Here the active traps evacuate quasiparticles efficiently in the superconducting leads of the main cooler.

At a lower bath temperature of 150 mK, the device behaves differently as shown in Fig. 3(b). At zero bias, the current of the active trap tends to overheat the superconducting leads. Electrons on the central normal island are subsequently heated up to 160 mK and 180 mK at $I_{\text{trap}}$ = 5 and 15 $\mu$A, respectively, at $V_C=0$. Nevertheless, at the optimum bias of the central cooler $V_{\text{C,opt}}$, the active trap improves the performance of the central cooler. Under finite values of $I_{\text{trap}}$, we reach $T_{\text{N,opt}}$ = 65 mK as compared to 80 mK when the active trap is off. These features result from the interplay of on chip heating, NIS cooling, and the temperature-dependent electron-phonon interaction in the various electrodes. The data on $T_{\text{N}}$ are collected as a function of $I_{\text{trap}}$ at different bath temperatures of 300, 200, and 90 mK in Figs. 3 (d), (e), and (f), respectively. We plot temperature differences $\Delta T_N = T_N(I_{\text{trap}}) - T_N(I_{\text{trap}}= 0)$ of the central island in two cases: when the main cooler is off ($V_C=0$, red circles), and under optimal bias ($V_{\text{C,opt}}$, blue squares). At the lowest bath temperature of 90 mK (Fig. 3 (f)), there is only heating at $V_C=0$ of the central cooler by non-zero $I_{\text{trap}}$. Still, the central cooler achieves a lower ultimate temperature of 65 mK at finite $I_{\text{trap}}$ as compared to 80 mK when the secondary cooler is off. 

A summary of our results is shown in Fig. 3 (c), where the optimum temperature drop on the central normal island, $\Delta T_{\text{N,opt}} = T_{\text{bath}} - T_{\text{N,opt}}$, is shown as a function of the bath temperature for various values of $I_{\text{trap}}$. At the low temperature end, the cooler reaches its base temperature of 65 mK from 200 mK, compared to 90 mK when $I_{\text{trap}}$ = 0. Around the bath temperature of 300 mK, we achieve the best performance in high-power SINIS cooler up to now: $\Delta T_{\text{N,opt}}$ = 200 mK, compared to $\Delta T_{\text{N,opt}}$ = 150 mK at $I_{\text{trap}}$ = 0 $\mu$A. Such a large temperature drop has previously been achieved only in devices with orders of magnitude lower cooling power \cite{PekolaPRL04}, where the overheating of the superconductor is easier to manage with passive trapping techniques.

\begin{figure}[t]
\begin{center}
\includegraphics[width=3in,keepaspectratio]{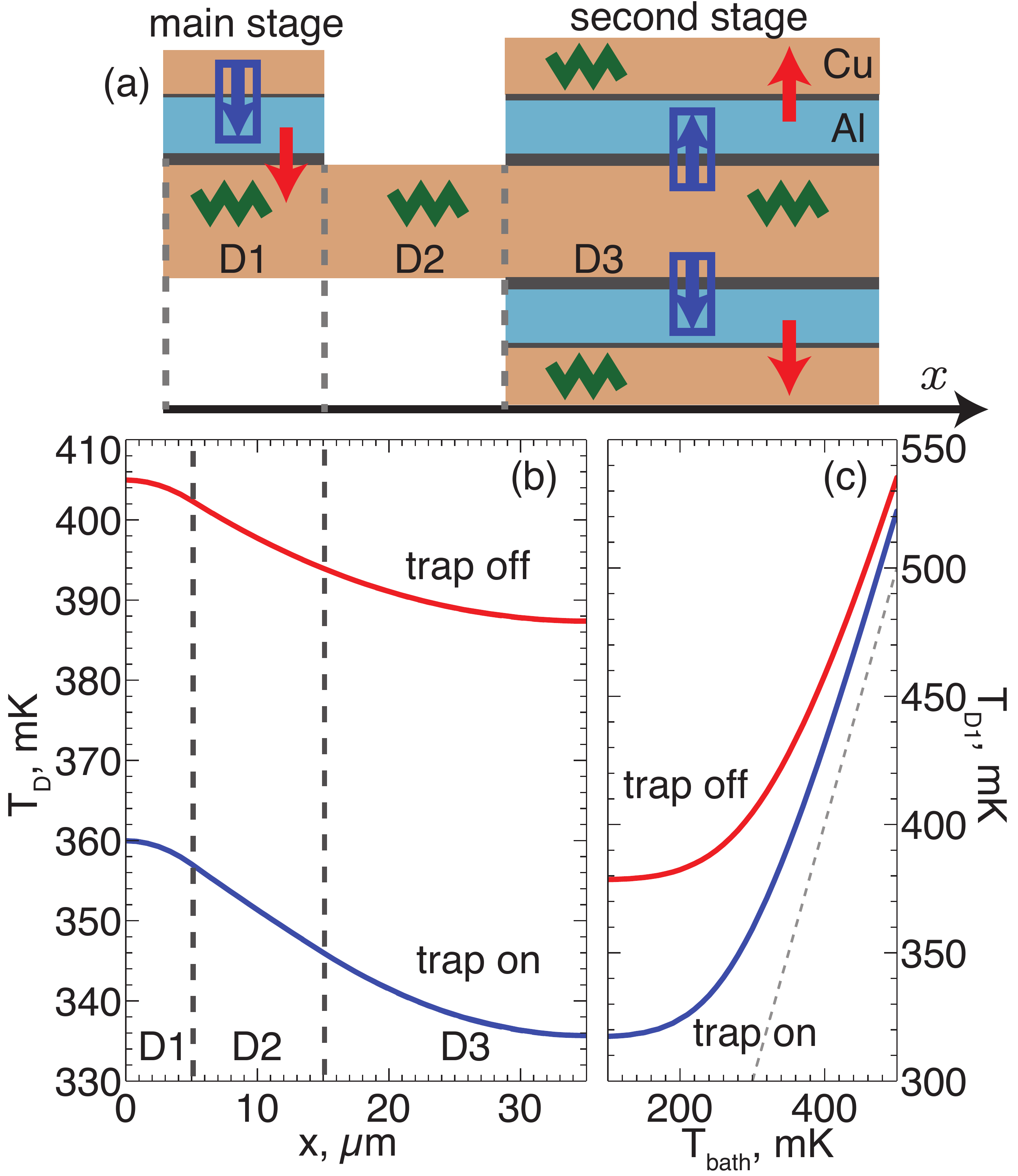}
\caption{(a) Thermal diagram of one half of the two-stage cooler where one NIS junction is assisted by one of the second stage coolers, the active trap. Color symbols represent main heat transport mechanisms: blue arrows for cooling power of a NIS junction, red arrows for dissipation of quasiparticles through a thin barrier, and green symbols for electron-phonon coupling in normal metal. (b) Temperature along the drain at bath temperature 300 mK based on the numerical thermal model for two cases: trap on and trap off. The drain is divided into three parts: D1 is directly underneath the main cooler, D3 is the active trap NIS junction, and D2 connects these two regions. (c) Temperature at the center of D1 as a function of bath temperature. The dashed line indicates $T_{D1}=T_{\text{bath}}$.}
\label{coolsol}
\end{center}
\end{figure}

\emph{Theoretical consideration:} Thermal transport of the device is illustrated by the diagram in Fig. 4 (a). Assume local electron temperatures at quasi-equilibrium throughout the device. Heat transport in each element is described by a one dimensional heat diffusion equation 
$\frac{\partial}{\partial x}\left(-\kappa_i\frac{\partial T_i}{\partial x}\right) = P_i$ with $T_i$ being the temperature, $\kappa_i$ the thermal conductance, and $P_i$ the total injected heat along the element $i$, which is either a superconductor (S), normal metal (N), or normal metal drain (D). The blue arrow denotes cooling by a NIS junction across a discrete barrier. Assuming a fully symmetric device, it extracts a power
\begin{eqnarray}
	\dot{Q}_{NIS}&=&\frac{1}{e^2R_N}\int{d\varepsilon(\varepsilon-\frac{eV}{2})n_S(\varepsilon)[f_N(\varepsilon-\frac{eV}{2})-f_S(\varepsilon)]}\nonumber\\
\end{eqnarray}
out of the normal metal and dumps a power IV/2+$\dot{Q}_{NIS}$ into the superconducting lead. Here, $f_N$ ($f_S$) is the distribution of electrons in the normal metal (superconductor). The red arrow represents heat transport through another NIS barrier to the drain at zero bias \cite{PeltonenPRB10}:
\begin{equation}
\dot{Q}_{V=0}=\frac{\sqrt{2\pi k_B\Delta^3}}{e^2 d_S r_D} (\sqrt{T_S}e^{-\Delta/k_BT_S}- \sqrt{T_D}e^{-\Delta/k_BT_D}),
\end{equation}
where $d_S$ is the thickness of the superconductor, and $r_D$ the resistivity of the tunnel barrier between the drain and the superconductor. The green symbol refers to the power exchanged between electrons and phonons \cite{WellstoodPRB}, 
\begin{equation}
	\dot{Q}_{e-\text{ph}}=\Sigma\mathcal{V}(T^5_e-T^5_{\text{ph}}),
\end{equation}
with $\Sigma$ the electron-phonon coupling coefficient and $\mathcal{V}$ the volume of the normal metal.

It is possible to write down a set of heat equations for each local temperature: $T_N$, $T_S$, $T_D$. This approach is similar to the thermal model that was presented in Ref. \cite{NguyenNJP}, where it matched well with experimental data. To illustrate the role of the second stage cooler, we solve for $T_D$ in the drain that connects to the active trap. As seen in Fig. 4 (a), the drain consists of three regions with different heat sources. Region D1, lying directly underneath the NIS junction, receives the secondary power of the main cooler, which is substantially larger than the cooling power itself due to the relatively low efficiency \cite{GiazottoRMP06}. The heat equation in this region writes
\begin{equation}
	\text{D1: } \kappa_N d_D \nabla^2T_{D1}=-\mathcal P_{e-\text{ph}}(T_{D1}) + IV/2.\nonumber
\end{equation}
This heat dissipates from region D1 to region D2, which connects the junction and the active trap. Here, electrons thermalize with phonons as
\begin{equation}
	\text{D2: } \kappa_N d_D \nabla^2T_{D2}=-\mathcal P_{e-\text{ph}}(T_{D2}). \nonumber
\end{equation}
In region D3, the heat is actively pumped away with the cooling power from the second stage cooler, on top of the electron phonon interaction as:
\begin{equation}
	\text{D3: } \kappa_N d_D \nabla^2T_{D2}=-\mathcal P_{e-\text{ph}}(T_{D2}) + \mathcal P_{NIS}(I_{\text{trap}},R_D), 
\end{equation}
Here, $d_D$ is the drain thickness, $\mathcal P_{e-\text{ph}}$ and $\mathcal{P}_{\text{NIS}}$ are local powers per unit area, corresponding to $\dot{Q}_{e-\text{ph}}$ and $\dot{Q}_{\text{NIS}}$. We solve this equation numerically for the bath temperature of 300 mK with parameters of the measured device: $R_N = 1000$ $\Omega\mu$m$^2$, $\Delta$ = 190 $\mu$eV, $d_D$ = 100 nm, the length of D1, D2, and D3 are 4, 10, and 20 $\mu$m, respectively. This result is shown in Fig. 4 (b) for two cases: when the trap is off and when it is optimally biased. Clearly, the active trap cools efficiently the drain, and the superconductor of the first stage. Figure 4 (c) shows the value of $T_D$ at the center of D1 region as a function of the bath temperature for the two cases, when the active trap is off and on, respectively. Above the bath temperature of 400 mK, the role of the active trap is diminished because of the overwhelming contribution from the electron-phonon coupling. 

To optimize the performance of a SINIS cooler, the superconducting electrode has to tolerate heat generated by the injection of quasiparticles. In accord with previous work \cite{ONeilPRB12,SukumarPRB12}, our series of experiments \cite{NguyenAPL,NguyenNJP,NguyenPRN} proves that the SINIS cooler reaches lower temperature with a better thermalized superconductor using a "passive" quasiparticle trap. Although the enhancement is impressive, the cooler does not meet theoretical expectations with this method only. Alternatively, the superconductor can be thermalized with a small perpendicular magnetic field \cite{PeltonenPRB10,MathieuNComm}. The field induces a small number of vortices with a normal core, which act as local quasiparticle traps. This method is not applicable for a large junction device, as the presence of a vortex also diminishes the quality of a superconductor. Closest to our work, Ferguson \cite{FergusonAPL} employs two aluminum films of different thickness, which act as two superconductors $S_1$ and $S_2$ with distinct gaps. A biased S$_1$IS$_2$ junction extracts efficiently excess quasiparticles out of a mesoscopic device, a Cooper pair transistor, located nearby on the same chip. 

The presented double stage cooler is the first experimental realization of a cascade tunnel junction refrigerator where one stage assists the operation of the other. Would one replace one of the stages with a superconductor of higher transition temperature $T_c$, e.g. tantalum \cite{Chaudhuri} or vanadium \cite{Quaranta}  as the second stage, the cascade cooler should be able to work starting from a higher bath temperature. Using titanium \cite{Faivre} as a superconductor with a lower $T_c$, the cooler should be able to reach a much lower $T_N$. Ultimately, employing two or three stages in cascade, it might achieve a temperature drop from 2 K to below 10 mK and replace a dilution cryostat in targeted applications \cite{LowellAPL13}. 

We highlight here a number of aspects that we have considered when designing the double stage refrigerator. First, instead of using AlMn \cite{NguyenNJP}, the quasiparticle drain is made of Cu, so that it is possible to connect it directly to the normal metal of the active trap. We found that this cooler performs quite similarly as the one with an AlMn drain, which is consistent with our expectations from the model in Ref.\cite{NguyenPRN}. Second, the NIS junctions of the active trap are clearly not optimized for lowering the temperature but to extract heat out of the main cooler. The junction area in the second stage is almost an order of magnitude bigger than that of the central cooler, so that its cooling power can compensate for the discarded heat from the central cooler. Last but not least, the active traps are positioned very close to the central cooler at a distance of 3 $\mu$m, to ensure an efficient evacuation of quasiparticles. If this distance is larger than the the electron-phonon relaxation length, which is about 10 $\mu$m at 300 mK for Cu \cite{NguyenPRN}, quasiparticles mostly lose heat to phonons and adversely increase the bath temperature of the central cooler, instead of being conveyed away by the active trap.

We have introduced an efficient method to thermalize non-equilibrium superconductors under active generation of quasiparticles. Powered by two SINIS coolers functioning as active quasiparticle traps, the proof-of-concept double stage refrigerator demonstrates a temperature drop from 300 mK to 100 mK at 300 pW cooling power. This serial design is promising, not only for electronic refrigeration, but it is also applicable for other cryo-electronic devices where one injects non-equilibrium quasiparticles into a superconductor. 

We thank H. Courtois, I. M. Khaymovich, and A. S. Mel'nikov for discussions. We acknowledge the support of the Academy of Finland through grants no. 284594 and 272218, and the Otaniemi Research Infrastructure for Micro and Nano Technologies (OtaNano). Samples were fabricated in the Micronova Nanofabrication Center of Aalto University.

\end{document}